\documentclass[12pt]{article}%
\usepackage{graphicx}
\usepackage{amsmath}
\usepackage{amsfonts}
\usepackage{amssymb}%
\providecommand{\U}[1]{\protect\rule{.1in}{.1in}}
\newtheorem{theorem}{Theorem}

\newtheorem{lemma}[theorem]{Lemma}

\newtheorem{proposition}[theorem]{Proposition}

\begin{document}

\title{\textbf{The characterization of ground states}}
\author{Jean Bellissard$^{1}$\thanks{Partially supported by NSF Grant 060096}\ \ ,
\ Charles Radin$^{2}$\thanks{Partially supported by Laboratoire d'Informatique
Fondamentale de Marseille and NSF Grant 0700120}\ \ , and \ Senya
Shlosman$^{3}$ \thanks{Partially supported by GREFI-MEFI and by the RFFI Grant
08-01-00105-a.}\\{\small $^{1}$ Georgia Institute of Technology, School of Mathematics, }\\{\small Atlanta GA 30332-0160, USA}\\{\small $^{2}$ Department of Mathematics, University of Texas, }\\{\small Austin, TX 78712-0257, USA}\\{\small $^{3}$ Centre de Physique Theorique, CNRS, }\\{\small Luminy Case 907, 13288 Marseille, Cedex 9, France }\\{\small and Institute for Information Transmission Problems }\\{\small (IITP RAS), Moscow, Russia} \vspace{0.2cm}}
\maketitle

\begin{abstract}
We consider limits of equilibrium distributions as temperature approaches
zero, for systems of infinitely many particles, and characterize the support
of the limiting distributions. Such results are known for particles with
positions on a fixed lattice; we extend these results to systems of particles
on $\mathbb{R}^{n}$, with restrictions on the interaction.

\end{abstract}

\section{Introduction}

Consider a physical system consisting of a large number of interacting
molecules in thermal equilibrium. Equilibrium statistical mechanics accurately
models such systems. The fundamental qualitative feature of having fluid and
solid phases, the latter appearing at low temperature and/or high pressure,
can even be usefully modelled with \textit{classical} statistical mechanics.
Yet although this phase structure has been amply supported by computer
simulations \cite{FS} there is as yet not a single model, of particles moving
in space and interacting through reasonable short range forces, in which such
fundamental features can be proven \cite{Ra1}. It is not difficult to model a
solid if one uses less physical interactions, as in the Einstein or Debye
models, but this has not yet been achieved with a satisfactory short range
model, which could also describe a fluid phase. (See however \cite{Ru1, LM,
Ra2}.) This is one of the main unsolved problems in condensed matter physics
\cite{Br, Uh}.

One of the difficulties is that to unambiguously characterize or distinguish a
phase one must use the thermodynamic limit, or, equivalently, uniformly
control the behavior of the system as the system size grows indefinitely. Now
consider the grand canonical ensemble for a system of finite size, for which
the unnormalized probability density of particle configuration $\omega$ is
$\exp-\beta\lbrack E(\omega)-\lambda|\omega|]$, where $E(\omega)$ is the
energy of $\omega$, $|\omega|$ is its particle number, $\lambda$ is the
chemical potential and $\beta$ is the inverse temperature. If one takes the
limit $\beta\rightarrow\infty$, for fixed $\lambda$, one easily sees that the
probability tends to concentrate on configurations $\omega$ which minimize
$E(\omega)-\lambda|\omega|$ and which are called ground state configurations.
This can be useful, as one can then understand the state at large $\beta$ as a
perturbation of the energy ground state, and try to understand the solid phase
from this. But, as we just noted, one must be careful with the order of
limits; one needs to see the approximation of positive temperature states by
zero temperature states \textit{uniformly in the size of the system}, and not
only is this not obvious, it can actually fail, as we show below.

In other words, a major difficulty in solving this old problem, of
successfully modelling the origin of the solid state in terms of short range
forces, is to control the approximation of low temperature states by those at
zero temperature (energy ground states) as the system size grows. One way to
systematize this is to make sense of infinite volume limit \textquotedblleft
Gibbs states\textquotedblright\ $\mu_{\beta,\lambda}$ for finite $\beta$,
introduce the \textquotedblleft ground states\textquotedblright\ $\mu
_{\infty,\lambda}$ as the limit points $\lim_{\beta\rightarrow\infty}%
\mu_{\beta,\lambda}$, and characterize the \textquotedblleft ground state
configurations\textquotedblright\ $\omega$ in the support of $\mu
_{\infty,\lambda}$, the smallest closed set of configurations of probability
1. This is the path we will take as it makes the control of the limit as
$\beta\rightarrow\infty$ a bit easier, having already taken the limit in the
size of the system. So $\mu_{\beta,\lambda}$ and $\mu_{\infty,\lambda}$ will
be probability distributions on configurations $\omega$ of particles in
unbounded physical space. From the above analysis in finite volume we expect
$\mu_{\infty,\lambda}$ to be supported by configurations $\omega$ which in
some sense minimize $E(\omega)-\lambda|\omega|$. Of course for an infinite
system $E(\omega)-\lambda|\omega|$ is typically going to be infinite, so one
must adjust appropriately both the defining characteristic of the equilibrium
distribution $\mu_{\beta,\lambda}$ and the optimization approached as
$\beta\rightarrow\infty$. For $\mu_{\beta,\lambda}$ this was solved rather
generally many years ago, and we use the result below. But for $\mu
_{\infty,\lambda}$ this was only solved in the simpler situation of models
with particles living in a discrete space, typically a lattice \cite{Ru2, Sc}.
Restricting oneself to lattice models is a weakness however if one is trying
to model the fluid/solid transition, or more specifically to model a solid.
Removing this obstacle is the main motivation for this work. We show that
under certain reasonable conditions the optimization characterization of the
support of $\mu_{\infty,\lambda}$ is the same for models of particles in space
as for models of particles on a lattice. The proof is somewhat harder, and for
good reason: it is known that without extra assumptions on the interaction the
particle density can be unbounded at any positive temperature, a phenomenon
not possible in typical lattice models.

Our arguments are necessarily technical since we are forced to deal carefully
with limits, but this is justified by the direct relevance of our results to
matters of importance to physical theory. We note that ground states
are also used significantly in optimization schemes outside physics; see for 
instance \cite{HR}.

\section{Convergence to Ground State Configurations}

First some notation and assumptions. We assume a two-body interaction
potential $U(s,t)$ dependent only on the separation of the point particles at
positions $s,t$ in $\mathbb{R}^{n}$, including a hard core at separation 1 and
diverging as the separation decreases to 1:%
\[
U(s,t)\left\{
\begin{array}
[c]{cc}%
=\infty & \text{ for }\left\vert s-t\right\vert \leq1,\\
\rightarrow\infty & \text{ for }\left\vert s-t\right\vert \searrow1.
\end{array}
\right.
\]
We assume $U$ has finite range $R>1$, and that $U\geq-m,$ for some $m>0$.
Denote the chemical potential by $\lambda$. 

We denote by $\Omega$ the set of all finite or countably infinite
configurations $\omega\subset\mathbb{R}^{n}$ of particles which are separated
by a distance at least $1$. By $\omega_{j}$ we denote the positions of the
particles in $\omega$, and by $b_{1}(\omega)$ the set of balls $b_{1}%
(\omega_{j})$ of diameter $1$ centered at positions $\omega_{j}$. For
$A\subset\mathbb{R}^{n}$ we denote by $\Omega_{A}$ the set of configurations
$\omega=\omega_{A}\equiv\omega\cap A$, which have all their particles in $A.$
The number of particles in $\omega_{A}$ will be denoted by $\left\vert
\omega_{A}\right\vert .$ With the usual topology $\Omega$ is compact.

Now we will introduce the notion of the Gibbs state. We first introduce for
every bounded $A$ and $\omega\in\Omega_{A}$ the energy:%
\begin{equation}
H(\omega)=\sum_{i<j}U(\omega_{i},\omega_{j})+\lambda\left\vert \omega
\right\vert .
\end{equation}
For two collections of particles, $\omega^{\prime}\in\Omega_{A},$
$\omega^{\prime\prime}\in\Omega,$ we define the interaction between them as%
\begin{equation}
H(\omega^{\prime},\omega^{\prime\prime})=\sum_{i,j}U(\omega_{i}^{\prime
},\omega_{j}^{\prime\prime}).
\end{equation}
and the sum
\begin{equation}
H(\omega^{\prime}{\LARGE |}\omega^{\prime\prime})=H(\omega^{\prime}%
)+H(\omega^{\prime},\omega^{\prime\prime}).
\end{equation}

We say that the probability measure $\mu_{\beta,\lambda}$ on $\Omega$ is a
Gibbs measure, corresponding to the interaction $U,$ inverse temperature
$\beta$ and chemical potential $\lambda,$ if for any finite $A$ and any
function $f$ on $\Omega_{A}$ we have%
\begin{align}
& \int f\left(  \omega_{A}\right)  \ d\mu_{\beta,\lambda}\left(
\omega\right)  \label{dlr}\\
& =\int\left[  \int f\left(  \omega_{A}\right)  \exp\left\{  -\beta
H(\omega_{A}{\LARGE |}\omega_{A^{c}})\right\}  \ d\pi_{A}\left(  \omega
_{A}\right)  \right]  Z_{\beta,\lambda}^{-1}\left(  A,\omega_{A^{c}}\right)
d\mu_{\beta,\lambda}\left(  \omega\right)  .\nonumber
\end{align}
Here 

\begin{itemize}
\item $A^{c}=\mathbb{R}^{n}\smallsetminus A,$  

\item $\pi_{A}$ is the Poisson measure on $\Omega_{A},$ which on the $k$-particle
subset of $\Omega_{A}$  is just the Lebesgue measure, normalized by the factor
$e^{-\left\vert A\right\vert }\frac{\left\vert A\right\vert ^{k}}{k!},$

\item  the \textit{partition function}
\[
Z_{\beta,\lambda}\left(  A,\omega_{A^{c}}\right)  =\int\exp\left\{  -\beta
H(\omega_{A}{\LARGE |}\omega_{A^{c}})\right\}  \ d\pi_{A}\left(  \omega
_{A}\right)  .
\]

\end{itemize}

It is easy to see that any such measure gives probability $1$ to the set of
configurations in which no two particles are at distance $1$.

The probability distribution $q_{A,\omega_{A^{c}}}\equiv q_{\beta
,\lambda,A,\omega_{A^{c}}}$ on $\Omega_{A},$ given by the density
$Z_{\beta,\lambda}^{-1}\left(  A,\omega_{A^{c}}\right)  \exp\left\{  -\beta
H(\omega_{A}{\LARGE |}\omega_{A^{c}})\right\}  $ with respect to the measure
$\pi_{A},$ is called the \textit{conditional Gibbs distribution},
corresponding to the boundary condition $\omega_{A^{c}}.$ The equation
$\left(  \ref{dlr}\right)  $ is called the Dobrushin-Lanford-Ruelle
(DLR) conditions; see \cite{D1,D2,D3,LR}. (The conditions are a way to
replace, for systems of infinite size, the usual formula which one uses for 
finite systems.) Any measure obtainable from such a Gibbs state by the limit 
$\beta \to \infty$ will be called a \textit{ground state}.

Let the set $G$ of \textit{ground state configurations}\ be defined as:%

\begin{align*}
G=\{ &  \omega\in\Omega:\text{ for every bounded }\Lambda\subset\mathbb{R}%
^{n}\text{ and every }\omega^{\prime}=\left(  \omega_{\Lambda}^{\prime}%
,\omega_{\Lambda^{c}}\right)  ,\text{ \ \ }\\
&  H(\omega_{\Lambda}^{\prime}|\omega_{\Lambda^{c}})-H\left(  \omega_{\Lambda
}|\omega_{\Lambda^{c}}\right)  \geq0\}.
\end{align*}

This set is nonempty, see \cite{Ra2}.

Our main result is the following

\begin{theorem}
\label{T1} Let $\mu_{\infty}$ be any limit point of the family of Gibbs states
$\mu_{\beta}$ as $\beta\rightarrow\infty,$ i.e. a ground state. Then
$\mu_{\infty}\left(  G\right)  =1.$
\end{theorem}

Theorem \ref{T1} holds in more a general situation, when the interaction has
no hard core, but possess instead the superstability property. (We
discuss superstability in the last section.) The proof is
more complicated, and we will not present it here. 

Theorem \ref{T1}  is equivalent to 

\begin{theorem}
\label{T2} Assume that $\omega\in G^{c}$. Then there exists an open
neighborhood $W$ of $\omega$ such that $\int_{W} ~d\mu_{_{\beta}}\left(
\sigma\right)  \rightarrow0$ as $\beta\rightarrow\infty$.
\end{theorem}

\textbf{Proof of Theorem \ref{T2}. }Before{ }giving the formal proof we
present its plan. If $\omega\in G^{c}$ then the following holds: there exists
a finite volume $B$, inside which the configuration $\omega\equiv\left(
\omega_{B},\omega_{B^{c}}\right)  $ can be modified into $\bar{\omega}%
\equiv\left(  \bar{\omega}_{B},\omega_{B^{c}}\right)  $ in such a way that
\begin{equation}
\Delta\left(  \omega\right)  \equiv H\left(  \omega_{B}{\LARGE |}\omega
_{B^{c}}\right)  -H\left(  \bar{\omega}_{B}{\LARGE |}\omega_{B^{c}}\right)
>0.
\end{equation}
We will be done if we can find open neighborhoods $W,\bar{W}$ of the
configurations $\omega$ and $\bar{\omega},$ such that
\begin{equation}
\frac{\mu_{\beta}\left(  W\right)  }{\mu_{\beta}\left(  \bar{W}\right)
}\rightarrow0
\end{equation}
as $\beta\rightarrow\infty.$ So we need to find an upper bound for $\mu
_{\beta}\left(  W\right)  $ and a lower bound for $\mu_{\beta}\left(  \bar
{W}\right)  .$ To do this we will use the following simple

\begin{lemma}
\label{L} For every value of the chemical potential $\lambda$ there exists a
distance $\rho\left(  \lambda\right)  >1$ such that the following holds for
all $\rho$ in the interval $\left(  1,\rho\left(  \lambda\right)  \right)  $:

Let $M\subset\mathbb{R}^{n}$ be any bounded volume and $\xi\in\Omega_{M^{c}}$
-- any \textquotedblleft boundary condition\textquotedblright. Denote by
$\Omega_{M,\rho}\left(  \xi\right)  \subset\Omega_{M}$ the subset
\begin{equation}
\left\{
\begin{array}
[c]{c}%
\sigma\in\Omega_{M}:\text{two particles of }\sigma\text{ are separated by
}<\rho,\text{ or a particle of }\sigma\text{ }\\
\text{ is at distance }<\rho\text{ from a particle of }\xi
\end{array}
\right\}  .
\end{equation}
Then the conditional Gibbs probability $q_{\beta,M,\xi}\left(  \Omega_{M,\rho
}\left(  \xi\right)  \right)  $ goes to $0$ as $\beta\rightarrow\infty.$ This
convergence, of course, is not uniform in $M,$ but for every $M$ it is uniform
in $\xi.$ {In other words,}
\begin{equation}
q_{\beta,M,\xi}\left(  \Omega_{M}\setminus\Omega_{M,\rho}\left(  \xi\right)
\right)  =1-\gamma\left(  \beta,M,\xi,\rho\right)  ,
\end{equation}
where for every $M,\rho$ the function $\gamma\left(  \beta,M,\xi,\rho\right)
\rightarrow0$ as $\beta\rightarrow\infty,$ uniformly in $\xi.$

The same statement holds for the subset
\begin{equation}
\Omega_{M,\rho}=\left\{  \sigma\in\Omega_{M}:\text{two particles of }%
\sigma\text{ are at distance }<\rho\right\}  ,
\end{equation}
since for every $\xi$ we have $\Omega_{M,\rho}\subset\Omega_{M,\rho}\left(
\xi\right)  .$
\end{lemma}

Without the hard core condition Lemma \ref{L} does not hold, and has to be
replaced by a weaker statement. Our proof of Lemma \ref{L} uses the divergence
of the repulsion near the hard core.

The proof of Theorem \ref{T2} proceeds now as follows. Let $\bar{B}$ be the
open $R$-neighborhood of $B$ in $\mathbb{R}^{n}.$ Due to the Lemma it is
enough to consider the case of $\omega$ such that $\omega_{\bar{B}}%
\not \in \Omega_{\bar{B},\rho(\lambda)}.$

By an $r$-perturbation of a finite configuration $\varpi\in\Omega$ we will
mean any finite configuration $\varkappa$ with the same number of particles,
such that for every particle $\varpi_{j}\in\varpi$ the intersection
$\varkappa\cap b_{1}(\varpi_{j})$ consists of precisely one particle
$\varkappa_{j}\in\varkappa,$ and $\mathrm{dist}(\varpi_{j},\varkappa_{j})<r.$

Now we define the open neighborhood $W$ of $\omega$ by putting
\begin{equation}
W=\left\{  \left(  \varkappa,\xi\right)  :\varkappa\in\Omega_{r}\left(
\omega,\bar{B}\right)  ,\xi\in\Omega_{\bar{B}^{c}}\right\}  ,
\end{equation}
where $\Omega_{r}\left(  \omega,\bar{B}\right)  $ is the set of all those
$r$-perturbations $\varkappa$ of $\omega_{\bar{B}}$ which also belong to
$\Omega_{\bar{B}}.$ It is immediate to see that if $r\leq\rho(\lambda)/2$ then
for every $\left(  \varkappa,\xi\right)  \in W$%
\begin{align*}
\left\vert H\left(  \varkappa_{B}\right)  -H\left(  \omega_{B}\right)
\right\vert  &  <Cr,\\
\left\vert H\left(  \varkappa_{B},\varkappa_{\bar{B}\setminus B}\right)
-H\left(  \omega_{B},\omega_{\bar{B}\setminus B}\right)  \right\vert  &  <Cr,
\end{align*}
for some $C=C\left(  B\right)  .$ Let $r$ be so small that $Cr<\frac
{\Delta\left(  \omega\right)  }{10}.$ Then, by DLR,%
\begin{align*}
&  \int_{W}d\mu_{_{\beta}}\left(  \varkappa,\xi\right)  \\
&  =\int_{\Omega_{r}\left(  \omega,\bar{B}\right)  }
\frac{\left[  \int_{\Omega_{r}\left(
\omega,\varkappa_{\bar{B}\setminus B}\right)  }\exp\left\{  -\beta H\left(
\varkappa_{B}{\LARGE |}\varkappa_{\bar{B}\setminus B}\right)  \right\}
d\pi_{B}(\varkappa_{B})\right]}{Z_{B}\left(
\varkappa_{\bar{B}\setminus B}\right)  }  d\mu_{_{\beta}}\left(  \varkappa,\xi\right)
\\
&  \leq\exp\left\{  -\beta\left[  H\left(  \omega_{B}{\LARGE |}\omega_{B^{c}%
}\right)  -\frac{\Delta\left(  \omega\right)  }{10}\right]  \right\}
\int_{\Omega_{r}\left(  \omega,\bar{B}\right)  }\frac{1}{Z_{B}\left(
\varkappa_{\bar{B}\setminus B}\right)  }d\mu_{_{\beta}}\left(  \varkappa
,\xi\right)  ,
\end{align*}
where
\begin{equation}
\Omega_{r}\left(  \omega,\varkappa_{\bar{B}\setminus B}\right)  =\left\{
\tilde{\varkappa}\in\Omega_{r}\left(  \omega,\bar{B}\right)  :\tilde
{\varkappa}_{\bar{B}\setminus B}=\varkappa_{\bar{B}\setminus B}\right\}  ,
\end{equation}

\noindent and $Z_{B}\left(  \varkappa_{\bar{B}\setminus B}\right)  $'s are the
{partition functions}.

In the same way, and recalling the meaning of $\bar{\omega}$, we put
\begin{equation}
\bar{W}=\left\{  \left(  \varkappa,\xi\right)  :\varkappa\in\Omega_{r}\left(
\bar{\omega},\bar{B}\right)  ,\xi\in\Omega_{\bar{B}^{c}}\right\}  .
\end{equation}
Without loss of generality we can assume that for the same $C$ and every
$\left(  \varkappa,\xi\right)  \in\bar{W}$%
\begin{align*}
\left\vert H\left(  \varkappa_{B}\right)  -H\left(  \bar{\omega}_{B}\right)
\right\vert  &  <Cr,\\
\left\vert H\left(  \varkappa_{B},\varkappa_{\bar{B}\setminus B}\right)
-H\left(  \bar{\omega}_{B},\omega_{\bar{B}\setminus B}\right)  \right\vert  &
<Cr.
\end{align*}
Then%
\begin{align*}
&  \int_{\bar{W}}d\mu_{_{\beta}}\left(  \varkappa,\xi\right)  \\
&  =\int_{\Omega_{r}\left(  \omega,\bar{B}\right)  }
\frac{\left[  \int_{\Omega_{r}\left(
\bar{\omega},\varkappa_{\bar{B}\setminus B}\right)  }\exp\left\{  -\beta
H\left(  \varkappa_{B}{\LARGE |}\varkappa_{\bar{B}\setminus B}\right)
\right\}  d\pi_{B}(\varkappa_{B})\right]}{Z_{V}\left(
\varkappa_{\bar{B}\setminus B}\right)  }  d\mu_{_{\beta}}\left(  \varkappa
,\xi\right)  \\
&  \geq\exp\left\{  -\beta\left[  H\left(  \bar{\omega}_{B}{\LARGE |}%
\omega_{B^{c}}\right)  +\frac{\Delta\left(  \omega\right)  }{10}\right]
\right\}  \int_{\Omega_{r}\left(  \omega,\bar{B}\right)  }
\hskip -.3truein\frac{\left[  \int_{\Omega
_{r}\left(  \bar{\omega},\varkappa_{\bar{B}\setminus B}\right)  }d\pi
_{B}(\varkappa_{B})\right]} {Z_{V}\left(  \varkappa_{\bar{B}\setminus B}\right)  }
 d\mu_{_{\beta}}\left(  \varkappa,\xi\right)  ,
\end{align*}

But the integral $\int_{\Omega_{r}\left(  \bar{\omega},\varkappa_{\bar
{B}\setminus B}\right)  }d\pi_{B}(\varkappa_{B})$ is just the Poisson measure
of the set $\Omega_{r}\left(  \bar{\omega}\right)  ,$ so it is a positive
number (not depending on $\beta$). The comparison of the last two estimates
proves our theorem. $\square$

\bigskip

\textbf{Proof of Lemma \ref{L}. }Using the above ideas it is straightforward.
Let $i\left(  n\right)  $ be the maximal number of particles with which any
given particle can interact. Suppose a particle $\varpi_{1}$ is $\rho$-close
to $\varpi_{2}.$ Due to the divergence of the repulsion near the hard core,
the energy of the interaction of particles $\varpi_{1}$ and $\varpi_{2}$
diverges as $\rho\searrow1,$ so we can assume that $U\left(  \varpi_{1}%
,\varpi_{2}\right)  >\lambda+i\left(  n\right)  m+1\ $once $\rho-1$ is small
enough. But then if we erase the particle $\varpi_{1},$ we gain at least one
unit of energy. The rest of the argument follows the same line as above.
$\square$

\section{Counterexamples}

In this section we will explain that some results which one might expect to
obtain in this area in fact do not hold.

Let $U$ be some pair potential, which is translation and rotation invariant,
i.e. $U\left(  s,t\right)  =U\left(  \left\vert s-t\right\vert \right)  .$ We
suppose $U$ to be superstable. Superstability is a property which means that
the repulsion part of the interaction dominates the attraction part, see
\cite{Ru3} for more details. The Lennard-Jones potential is an example of such
an interaction. Let $\lambda$ be some chemical potential.

Our initial modest goal {was to prove that} for all reasonable interactions
$U$ the following{ holds: }

\noindent \textbf{Statement.}\textit{ Let }$\beta_{n}\rightarrow\infty$\textit{ be a
sequence of inverse temperatures, going to infinity, and let }$\mu_{n}%
$\textit{ be a weakly converging sequence of Gibbs states, corresponding to
the interaction }$U,$\textit{ chemical potential }$\lambda$\textit{ and
inverse temperatures }$\beta_{n},$\textit{ i.e. }$\mu_{n}\in\mathcal{G}\left(
U,\lambda,\beta_{n}\right)  .$\textit{ Then the limiting state }$\mu_{\infty}%
$\textit{ is supported by the set }$G$\textit{ of ground state
configurations.}

\textit{ For every }$U,$\textit{ }$\lambda$\textit{ there exist a pair of
constants }$\underline{R}<\overline{R},$\textit{ such that for every }%
$\omega=\{\omega_{k}\}\in G$\textit{ }%
\begin{equation}
\inf_{i<j}\left\vert \omega_{i}-\omega_{j}\right\vert >\underline
{R},\label{21}%
\end{equation}
\textit{and every ball }$B_{r}\subset R^{n}$\textit{ of radius }%
$r>\overline{R}$\textit{ contains at least one particle }$\omega_{i}\in
\omega.$

\bigskip

If true, these properties would be a zero-level approximation to the ordered
structure which is expected (in some sense) to be formed by
solids. (For the use of these Delone properties in modelling the
ground states of
quasicrystals see \cite{BHZ,BL,Ke,La,Ra2,RW,VGW}.)

We have to say that we do not expect the above picture to hold without extra
assumptions, though these assumptions are expected to be mild and physically
natural. The following results point to difficulties that must be overcome.

Let $\mu\in\mathcal{G}\left(  U,\lambda,\beta\right)  $ be a Gibbs field with
inverse temperature $\beta,$ and the superstable interaction has attractive
part. Denote by $\rho_{\mu}\left(  x\right)  $ the expected number of
particles of the field $\mu$ in the unit ball centered at the point
$x\in\mathbb{R}^{n}.$

\begin{proposition}
\label{p1} {For every }$\lambda,\beta${ and }$U${ without hard core} there
exists a state $\bar{\mu}\in\mathcal{G}\left(  U,\lambda,\beta\right)  ,$ such
that the function $\rho_{\bar{\mu}}\left(  \cdot\right)  $ is unbounded on
$\mathbb{R}^{n}.$
\end{proposition}

{That statement means that the relation }$\left(  \ref{21}\right)  ${ cannot
hold in general.}

\begin{proposition}
\label{p2} Suppose that the state $\tilde{\mu}\in\mathcal{G}\left(
U,\lambda,\beta\right)  $ has the density function $\rho_{\tilde{\mu}}\left(
\cdot\right)  ,$ which is polynomially bounded, i.e. there exists a polynomial
$P\left(  \cdot\right)  ,$ such that $\rho_{\tilde{\mu}}\left(  x\right)  \leq
P\left(  x\right)  ,$ $x\in\mathbb{R}^{n}.$ Then there exists a constant
$C=C\left(  U,\lambda\right)  ,$ such that $\rho_{\tilde{\mu}}\left(
x\right)  \leq C.$
\end{proposition}

The proof of Proposition \ref{p2} can be obtained by the application of the
technique of compact functions, developed by R. Dobrushin in \cite{D4},
 see also \cite{DP}. Being proven,
Proposition \ref{p2} can be used to deduce the existence of the constants
$\underline{R},\overline{R}$ above, under condition that the random fields we
are dealing with have their density functions polynomially bounded (and hence,
uniformly bounded).

Proposition \ref{p1} can be derived from Proposition \ref{p2} and the
following construction. We will consider the 1D case; the generalization to
higher dimensions is obvious. Let us suppose that $U\left(  r\right)  \geq0$
for $r<r_{1},$ $U\left(  r\right)  <0$ for $r_{1}<r<r_{2},$ and $r_{1}<1/3,$
$r_{2}>2.$ Let $I_{n}$ be the unit segment centered at the integer point
$n\in\mathbb{R}^{1}.$ Let $\varpi^{-1},$ $\varpi^{1}$ be two configurations in
the segments $I_{-1},I_{1},$ and consider the conditional Gibbs distribution
$q\left(  \omega^{0}{\LARGE |}\varpi^{-1},\varpi^{1}\right)  $ in $I_{0},$
given configuration $\varpi^{-1}\cup$ $\varpi^{1}$ outside. Let $K>0$ be
fixed. Clearly, there exists a number $N\left(  1\right)  ,$ such that if
$\left\vert \varpi^{-1}\right\vert >N\left(  1\right)  ,$ $\left\vert
\varpi^{1}\right\vert >N\left(  1\right)  ,$ then $2K>\mathbb{E}\left(
\left\vert \omega^{0}\right\vert \right)  >K.$ Indeed, there will be a part of
the segment $I_{0},$ where the potential defined by the particles $\varpi
^{-1}\cup$ $\varpi^{1}$ will be attractive ($\equiv$negative), and the more
particles we will have in $\varpi^{-1}\cup$ $\varpi^{1},$ the deeper this well
will be.   Let now $\varpi^{-2},$ $\varpi^{2}$ be two configurations in the
segments $I_{-2},I_{2},$ and consider the conditional Gibbs distribution
$q\left(  \omega{\LARGE |}\varpi^{-2},\varpi^{2}\right)  $ in $I_{-1}\cup
I_{0}\cup I_{1},$ given configuration $\varpi^{-2}\cup$ $\varpi^{2}$ outside.
Clearly, there exists a number $N\left(  2\right)  ,$ such that if $\left\vert
\varpi^{-2}\right\vert >N\left(  2\right)  ,$ $\left\vert \varpi
^{2}\right\vert >N\left(  2\right)  ,$ then $\mathbb{E}\left(  \left\vert
\omega^{1}\right\vert \right)  >N\left(  1\right)  ,$ $\mathbb{E}\left(
\left\vert \omega^{-1}\right\vert \right)  >N\left(  1\right)  ,$ and so again
$\mathbb{E}\left(  \left\vert \omega^{0}\right\vert \right)  >K.$ Here we
denote by $\omega^{k}$ the restriction of $\omega$ on the segment $I_{k}.$ If
the number $N\left(  2\right)  $ is not too big, then we have in addition that
$\mathbb{E}\left(  \left\vert \omega^{0}\right\vert \right)  <2K.$ We can
repeat this construction inductively in $n.$ As a result, by taking a limit
point we get an infinite volume Gibbs state on $\mathbb{R}^{1},$ such that
$\mathbb{E}\left(  \left\vert \omega^{0}\right\vert \right)  >K.$ If $K$ is
chosen large enough: $K>C\left(  U,\lambda\right)  ,$ then the so constructed
state has the function $\rho\left(  \cdot\right)  $ unbounded, due to
Proposition \ref{p2}.

In the previous section we presented a {proof of our Statement}, restricted to
the case of interactions with hard core. This assumption plays a technical
role, and with an extra effort it can be removed.

\end{document}